\documentclass[conference,a4paper]{IEEEtran}
\usepackage{xcolor}
\usepackage{balance}
\usepackage[section]{placeins}
\usepackage{amsmath}
\usepackage{mwe}
\usepackage{subcaption}
\usepackage{graphicx}
\usepackage{lipsum}
\usepackage{adjustbox}

\ifCLASSOPTIONcompsoc
 \usepackage[caption=false,font=normalsize,labelfont=sf,textfont=sf]{subfig}
\else
 \usepackage[caption=false,font=footnotesize]{subfig}
\fi
\begin{document}
%
\title{A Metasurface-based Cross-slot Aperture Coupled Antenna Array for Ku-band Satellites in Reception}


\author{\IEEEauthorblockN{
Praveen Naidu Vummadisetty\IEEEauthorrefmark{1}, Aral Ertug Zorkun\IEEEauthorrefmark{1}, Juan Andr\'es V\'asquez Peralvo \IEEEauthorrefmark{1},
Mehmet Abbak\IEEEauthorrefmark{2}, 
 \\
Eva Lagunas\IEEEauthorrefmark{1}, 
Symeon Chatzinotas\IEEEauthorrefmark{1}   
}                                     
\IEEEauthorblockA{\IEEEauthorrefmark{1}
Interdisciplinary Centre for Security Reliability and Trust, University of Luxembourg, 1855 Luxembourg-Luxembourg, \\ (e-mails: \{praveen.vummadisetty, aral.zorkun, juan.vasquez, eva.lagunas, symeon.chatzinotas\}@uni.lu).}

\IEEEauthorblockA{\IEEEauthorrefmark{2}
SES, Chateau de Betzdorf, L-6815, Luxembourg, \\ (e-mails:\ abbak@ses.com).}

}



\maketitle

\begin{abstract}
The demand for cost-effective, low-profile user terminals for satellite communications supporting multicast services for Geostationary Orbit (GEO) satellites, has become a key focus for many Direct-to-Home (DTH) providers where the high data rates in the downlink are required. Planar antenna arrays with increased frequency bandwidth and improved ratio using meta-surfaces are considered as an effective solution for such systems. This paper presents a low-cost, aperture-coupled metasurface-enhanced patch antenna, operating within the 10.7-12.7 GHz frequency range. The antenna is designed to achieve a realized gain of at least 27 dBi across the band of interest using 32 x 32 array antennas distributed in a rectangular lattice. Initially configured for linear polarization, the antenna can be upgraded to support dual or circular polarization if required.
\end{abstract}

\vskip0.5\baselineskip

%

\section{Introduction}
The demand for high-speed data access or transmission is exponentially growing with the increasing number of end users, especially, in the Satellite Communication (SATCOM) systems. In recent years, the development of Geostationary Orbit (GEO) satellite systems has gained great importance in expanding communication coverage with increased data transfer rate \cite{chen_phased_2022}. However, in millimeter and sub-millimeter Radio Frequency (RF) bands signal transmission endures severe effects that decrease overall Quality of Service (QoS), such as high path losses, low beam directivity, and high interference, in GEO satellite systems. Therefore, developing novel technologies and techniques is essential to mitigate the drawbacks of signal transmission in high RF frequencies. 

The antenna arrays are essential in increasing the overall QoS in GEO satellite systems. The general description of antenna array is assembling a group of primarily identical antenna elements. To increase system performance in terms of increased gain, higher interference suppression, reduced back lobe levels, and increased beam directivity across a wide coverage area, antenna array technology combines multiple individual antenna elements \cite{chen_phased_2022}. 

The antenna elements' radiation patterns and electrical characteristics are the key factors determining antenna arrays' overall performance. Therefore, designing an antenna element per the application requirements is crucial. In millimeter and sub-millimeter RF frequency bands, the traditional antenna design methods might fail to meet the high gain and beam directivity, wide bandwidth, and lower side-lobe level requirements \cite{chen_phased_2022,milias_metamaterial-inspired_2021,tadesse_application_2020}. Therefore, metasurface-based antennas have been widely used to increase the overall performance and reduce the volume and active device cost of the antenna arrays in satellite systems \cite{chen_phased_2022,babaeian_high_2018}. Design of low-profile and low-cost antennas are essential \cite{aliqab_design_2023,tadesse_application_2020}. The use of artificial intelligence for the design has also been proposed and could alleviate the design timing \cite{fontanesi2023artificial}. 

This paper proposes a low-profile cross-slotted aperture coupled meta surface enhanced microstrip patch antenna operating in the 10.7-12.7 GHz frequency band with linear polarization. The aperture-coupled feeding lines on the ground plane are located right underneath and parallel to the cross-slot at the center of the microstrip patch antenna. The cross-slot aperture feeds the microstrip patch antenna and allows dual polarization when the feeding network is oriented vertically or horizontally. An additional metasurface layer is placed just above and parallel to the microstrip patch antenna to increase the bandwidth. A 4x4 array of Sieven-Piper \cite{clavijo_design_2003} structure without via is adopted for the meta-surface, and the radiating meta-surface can be considered a periodic structure. Then, the proposed structure is extended to a 32x32 antenna array configuration with an overall size of 42 cm x 42 cm. The proposed meta-surface enhanced the bandwidth of the antenna structure and array. The full-wave EM simulation results are obtained using CST.

In Section \ref{UC}, a brief introduction to unit cell selection is explained. In Section \ref{Design}, the design strategy of the proposed meta-surface enhanced antenna and the antenna array are given. In Section \ref{Sim} the simulation results of the proposed antenna array are given. Finally, Section \ref{Conc} concludes the paper. 

\section{Unit Cell Selection} \label{UC}
Various parameters affect the characteristics and overall performance of antenna arrays. The main factors affecting the antenna array's overall performance are the unit cell`s radiation pattern, directivity, gain, etc. Therefore, designing a unit cell according to the application requirements is crucial.

\begin{figure} [!htbp]  
\centering
\includegraphics[width=0.8\columnwidth] {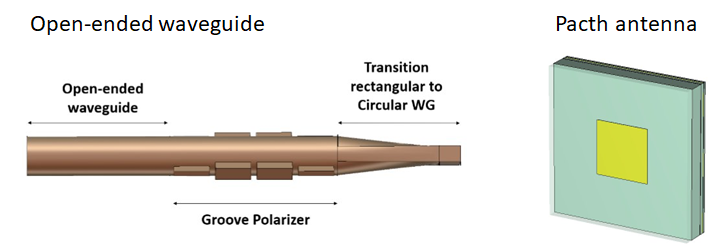}
\caption{The open-ended wave-guide \cite{vasquez-peralvo_flexible_2023} and patch antenna illustration for the unit cell.}
\label{fig:OEP}
\end{figure}

In antenna arrays, unit cells can be horn antennas \cite{tebbe_simulation_2016}, open-ended wave-guides \cite{vasquez-peralvo_flexible_2023}, or planar patches \cite{liu_metamaterial-based_2015}.  In Fig. \ref{fig:OEP}, illustrations of an open-ended wave-guide and patch unit cells are given.

The horn antennas or open-ended waveguides can be effective solutions since they provide low losses, high power capabilities, good heat dissipation, and broad bandwidth. However, horn antennas and open-ended waveguides are expensive, have high cross-polarization, and feeding networks for such arrays are complex to integrate. On the other hand, planar patch antennas allow multi-polarization, are easy to manufacture and integrate, and are cheap and lightweight at the cost of high losses and limited bandwidth.

Since the aim is to design dual-linearly polarized, low-profile, and low-cost antenna solutions for GEO satellite communication, planar patch antennas are the best option.

\section{Design}
This section presents the design procedure for the proposed unit cell antenna structure.

\subsection{Substrate Stack-up and Dimensions}
\begin{figure} [!htbp]  
\centering
\includegraphics[width=0.8\columnwidth] {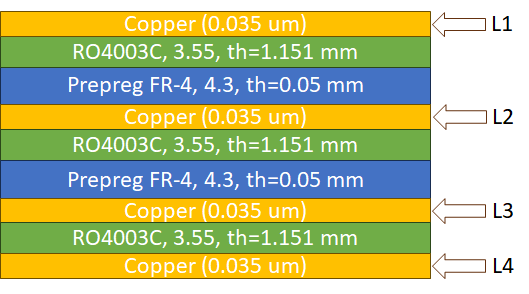}
\caption{Substrate stack-up.}
\label{fig:Stackup}
\end{figure}

\begin{figure} [!htbp]  
        \centering
        \begin{subfigure} [b]{0.475\columnwidth}  
            \centering
            \includegraphics[width=\linewidth]{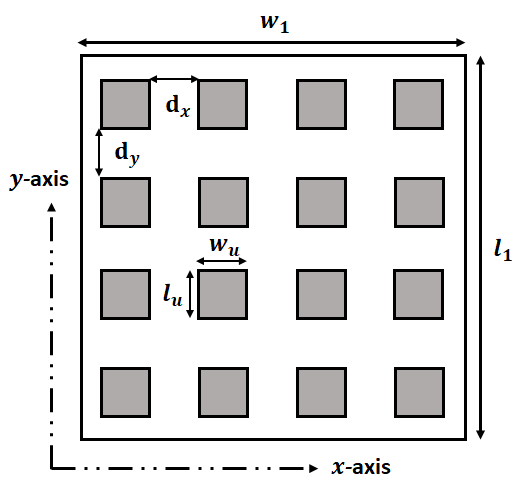}
            \caption[Network2]%
            {{\small}}    
            \label{fig: L1}
        \end{subfigure}
        \hfill
        \begin{subfigure} [b]{0.475\columnwidth}  
            \centering 
            \includegraphics[width=\linewidth]{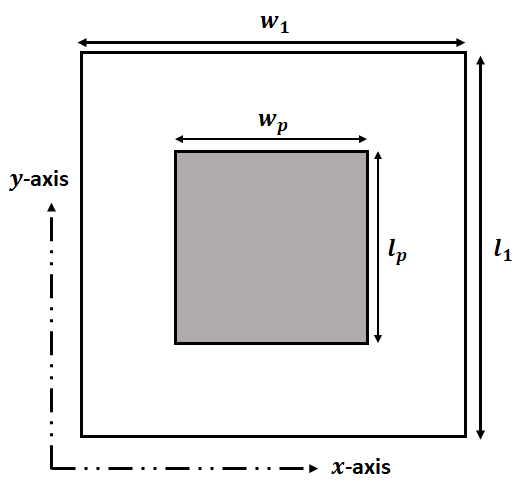}
            \caption[]%
            {{\small}}    
            \label{fig: L2}
        \end{subfigure}
        \vskip\baselineskip
        \begin{subfigure}  [b]{0.475\columnwidth}  
            \centering 
            \includegraphics[width=\linewidth]{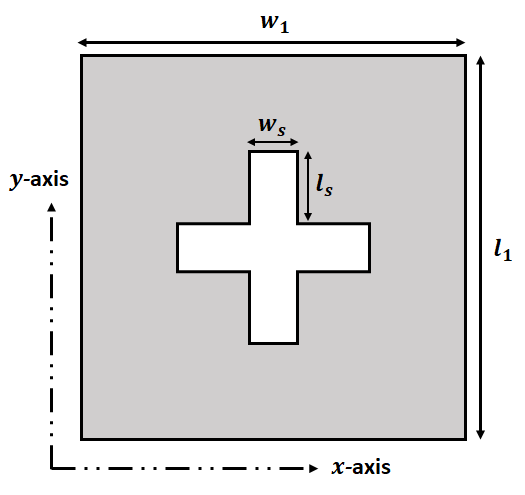}
            \caption[]%
            {{\small}}    
            \label{fig: L3}
        \end{subfigure}
        \hfill
        \begin{subfigure} [b]{0.475\columnwidth}  
            \centering 
            \includegraphics[width=\linewidth]{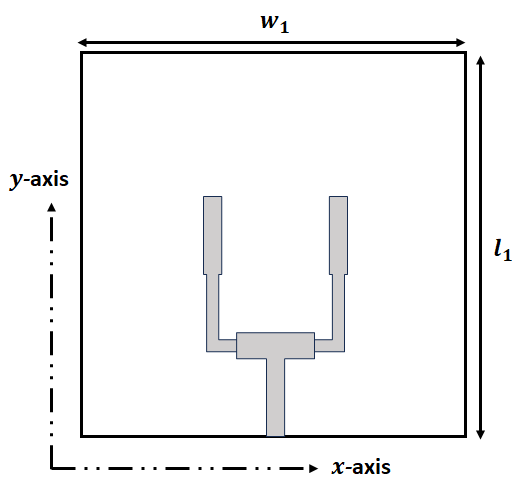}
            \caption[]%
            {{\small}}    
            \label{fig: L4}
        \end{subfigure}
        \caption[ Layout of the proposed unit cell structure.]
        {\small Layout of the proposed unit cell structure, a) L1: 4x4 array Sieven-Piper structure, b) L2: radiating patch antenna element, c) L3: cross-slotted ground plane, d) L4: feeding network.} 
        \label{fig:Layers}
    \end{figure}

Fig. \ref{fig:Stackup} presents the substrate stack-up of the unit cell, and the array consists of the proposed unit cell. Four conductor layers are utilized for the unit cell to fit the whole structure in the same footprint. As Fig. \ref{fig:Stackup} refers, four conductor layers are named and utilized as follows: L1 is used for the meta-surface (4x4 array of Sieven-Piper structure), L2 is used for radiating patch antenna element, L3 is used for cross-slotted ground plane, and L4 is used for split feeding network. The core dielectric substrates are Rogers RO4003C, and the prepreg layers are FR-4 substrates. Fig. \ref{fig:Layers} demonstrates the layers of the proposed unit cell structure, and Table \ref{Table1:Dimensions} shows the dimensions of the components.   

\begin{table} [ht]
\renewcommand{\arraystretch}{1.3}
\caption{Dimensions of The Proposed Structure.}
\label{Table1:Dimensions}
\centering

\begin{tabular}{|c||c||c||c||c|}
\hline
$w_{1}$ & $l_{1}$ & $d_{x}$ & $d_{y}$ & $w_{u}$\\
\hline
12.87 mm & 12.87 mm & 0.40 mm & 0.40 mm & 2.37 mm\\
\hline
$l_{u}$ & $w_{s}$ & $l_{s}$ & $w_{p}$ & $l_{p}$\\
\hline
2.37 mm & 2.04 mm & 0.15 mm & 5.04 mm & 5.04 mm\\
\hline
\end{tabular}
\end{table}

\subsection{Unit Cell Design} \label{Design}
\subsubsection{Aperture Design} \label{ApDesign}
In this study, we adopted aperture coupling to indirectly excite the patch antenna, where the energy from the transmission line is coupled through a slot in the ground layer to the patch. This technique not only increases the overall bandwidth of the antenna system but also eliminates the use of via where manufacturing tolerances are limited. The aperture of the patch antenna coupled with a transmission line can be estimated as:

\paragraph*{Step 1: Calculate the Patch Dimensions}

The formula can approximate the resonant length \( L \) of the patch:

\begin{equation}\label{equ:ResFreq}
L \approx \frac{c}{2f_0 \sqrt{\epsilon_{\text{eff}}}} .
\end{equation}
where \( c \) is the speed of light in vacuum (\(3 \times 10^8\) m/s),
\( f_0 \) is the resonant frequency,  and \( \epsilon_{\text{eff}} \) is the effective dielectric constant of the substrate.
\paragraph*{Step 2: Aperture Dimensions}

The length and width of the aperture are typically chosen to be a fraction of the patch's dimensions as:
\begin{equation}\label{equ:LW}
\begin{split}
a_{\text{L}} \approx \frac{L}{10} 
\\
a_{\text{W}} \approx \frac{W}{10}
\end{split}
\end{equation}
where \( W \) is the width of the patch.

\subsubsection{Patch Antenna Design}
\begin{figure} [!htbp]  
\centering
\includegraphics[width=0.8\columnwidth]{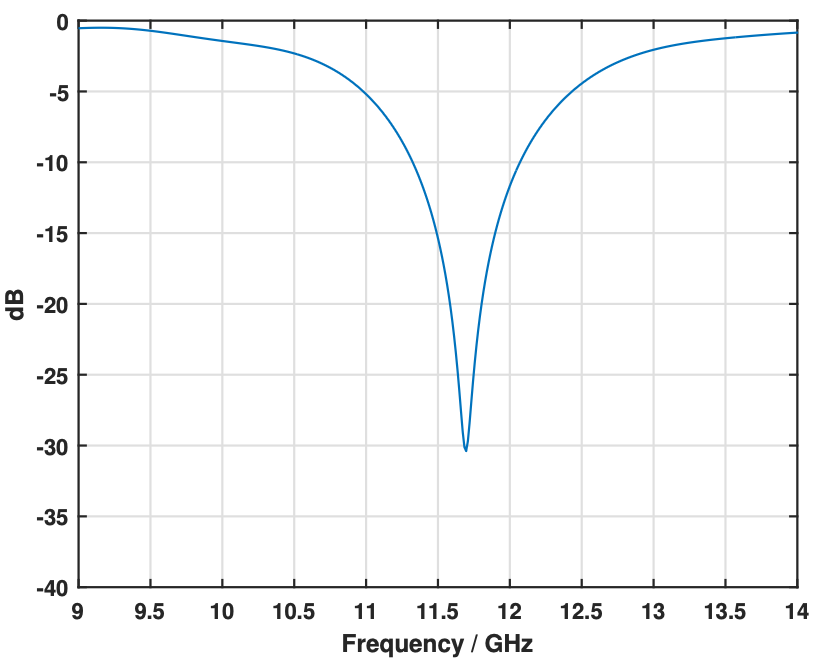}
\caption{Simulation results (S-11 parameters) of the cross aperture, coupled patch antenna.}
\label{fig:UCDNP}
\end{figure}

The design of the patch antenna was initially analyzed in the
previous subsection, as referenced in (\ref{equ:ResFreq}). To complement this design, a substrate
was selected that achieves an optimal balance between bandwidth enhancement and minimization
of surface waves. Additionally, to reduce costs, we opted to use the same substrate
type that was employed in the distribution network stage.
Following the optimization process, we determined that a substrate thickness of approximately
1.5 mm would be ideal. This has been rounded to 1.524 mm, a standard thickness that is
commercially available, ensuring ease of procurement and consistency in manufacturing. The
detailed design of the patch, along with the simulation results, are summarized in Fig. \ref{fig:UCDNP}.
\vspace{-0.5cm}

\subsubsection{Meta-surface Design}
\begin{figure} [!htbp]  
\centering
\includegraphics[width=0.8\columnwidth]
{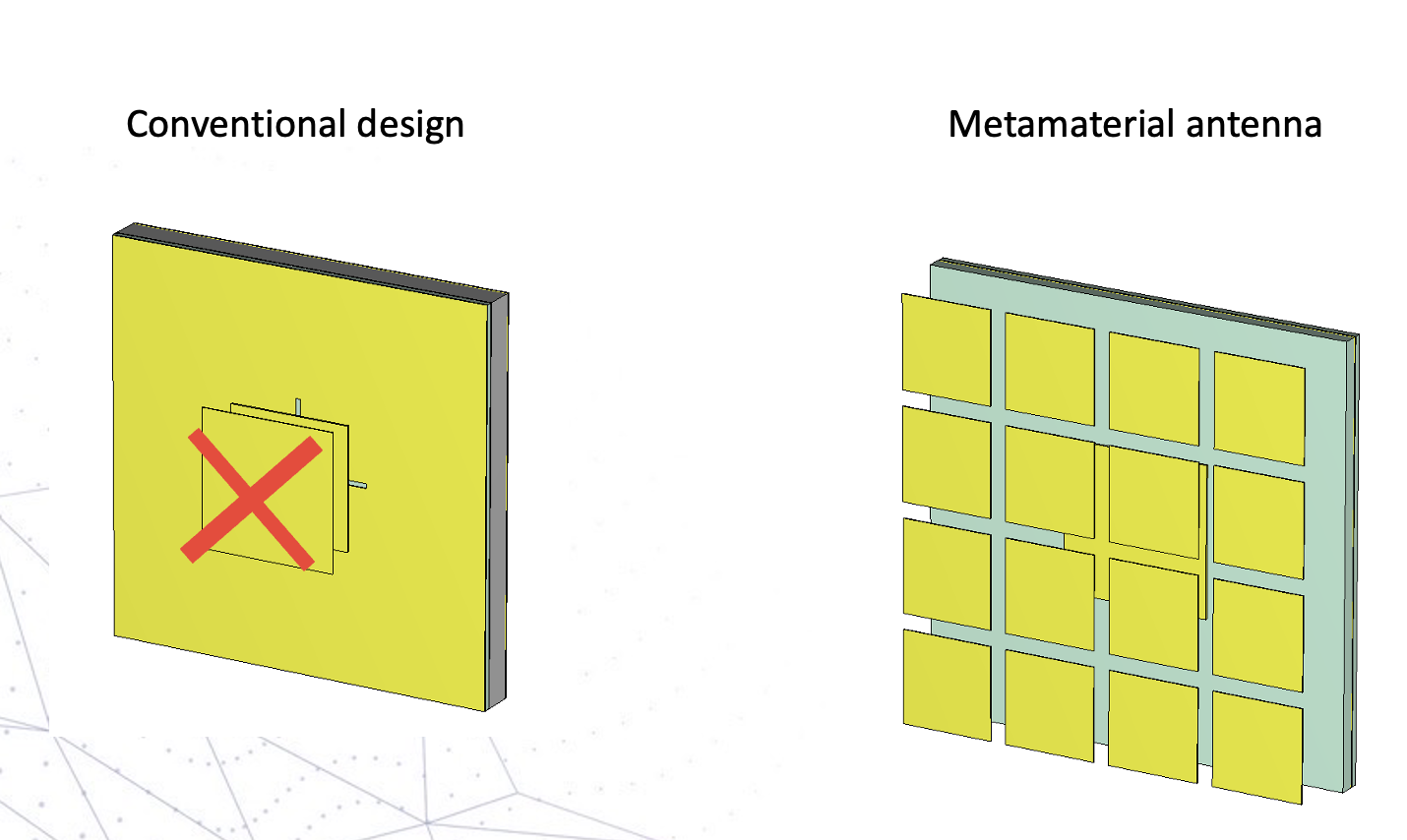}
\caption{Visually representation of the metasurface concept as the radiating element.}
\label{fig:META}
\end{figure}

\begin{figure} [!htbp]  
\centering
\includegraphics[width=0.8\columnwidth]
{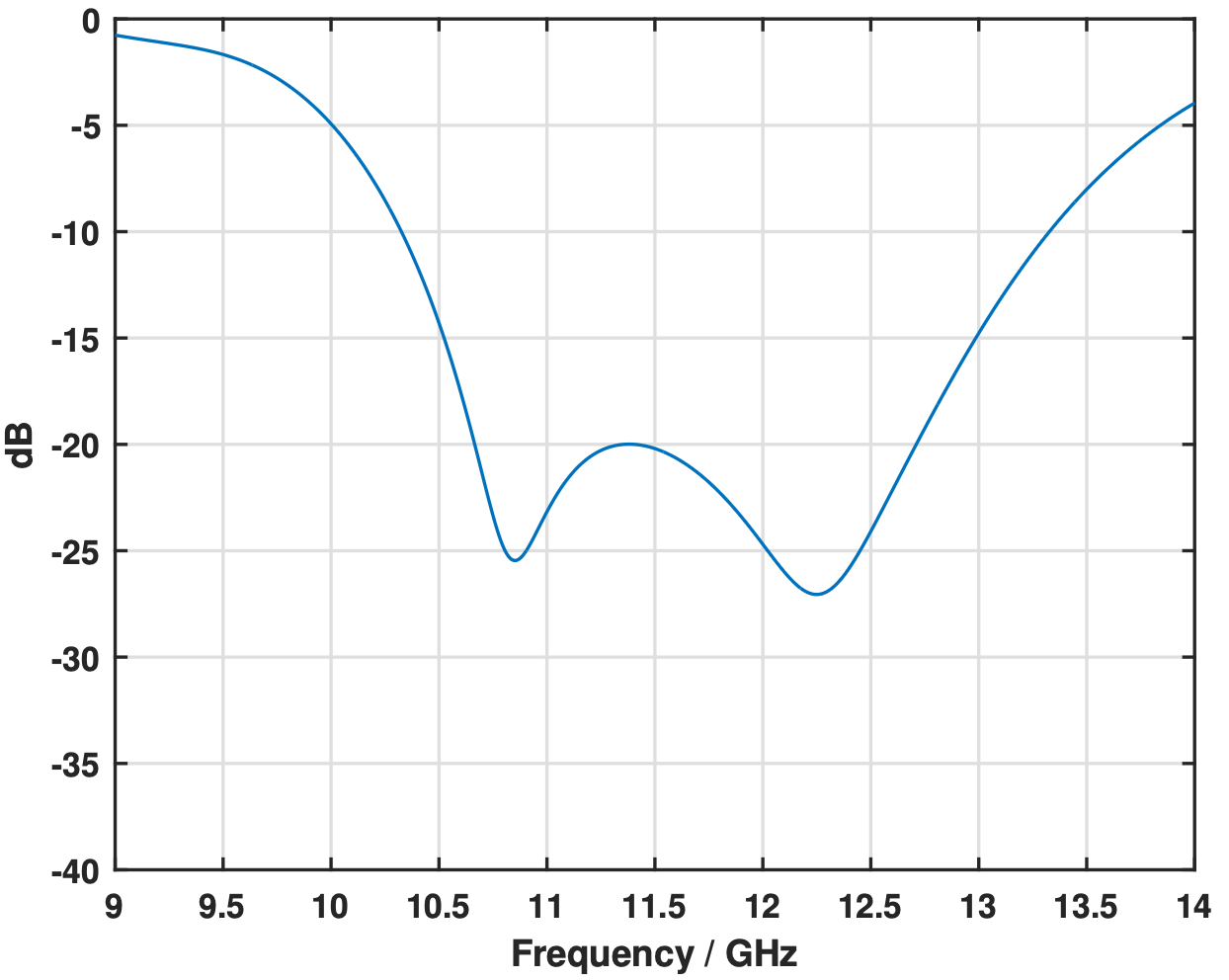}
\caption{Proposed metasurface and its frequency (S-11 parameters) response.}
\label{fig:OptMeta}
\end{figure}

We added a metasurface layer as an additional radiation element to increase the bandwidth. The meta-surface is strategically positioned over the previously designed patch antenna to optimize performance. This innovative configuration is visually depicted in Fig. \ref{fig:META}, illustrating the integration and functional synergy between the metasurface
and the underlying patch antenna.
The design process for the metasurface involved implementing a 4x4 array of Sievenpiper
structures without vias within the regions of the previously designed unit cell. The dimensions
of the patches and the separation between elements were refined through a final optimization
process, which will be detailed in the subsequent subsection. To comprehend the impact of the
metasurface, we conducted an analysis using Floquet ports, from which we derived the 
parameter, the dispersion diagram, and the current distribution.
The results of this analysis, depicted in Fig. \ref{fig:OptMeta}, demonstrate that the unit cell achieves
a high impedance crossing at approximately 10 GHz, indicating a substantial bandwidth and
enabling the transmission of Mode 1 within the intended frequency range. Additionally, the
current distribution is predominantly concentrated at the edges of the patches, suggesting that
the structure effectively manages a considerable bandwidth.
\vspace{-0.4cm}

\subsubsection{Global Optimization}
To meet all specified requirements for the antenna design, a comprehensive global optimization was conducted on the unit cell components. Below, we outline each component along with its associated optimization parameters:

\begin{itemize}
    \item \textbf{Metasurface}: Inter-element spacing and Sievenpiper patch length.
    \item \textbf{Patch antenna}: Length of the patch.
    \item \textbf{Aperture}: Length and width of the aperture.
    \item \textbf{Substrate thickness}: Thickness of the substrates used.
    \item \textbf{Distribution network}: Width and length of each transmission line.
\end{itemize}

The objectives of this optimization include:
\begin{itemize}
    \item Achieving an \(S_{11}\) of -20 dB within the frequency range of 10.7 - 12.7 GHz.
    \item Maintaining a high front-to-back ratio.
    \item Ensuring an efficiency greater than 95\%.
    \item Achieving a stable and increasing gain trend as the frequency increases.
\end{itemize}
In addition, the unit cell has been simulated with periodic boundary conditions to emulate the effect of mutual coupling with neighbour unit cells. This was configured before the optimization process.

The optimization utilized the trusted region framework algorithm and required approximately 10 hours on a standard desktop computer (Core i7, 32 GB of RAM) to meet the aforementioned conditions. The results of this optimization process are visually depicted in Figure \ref{fig:Optimization}.

\begin{figure} [!htbp]  
\centering
\includegraphics[width=0.8\columnwidth]
{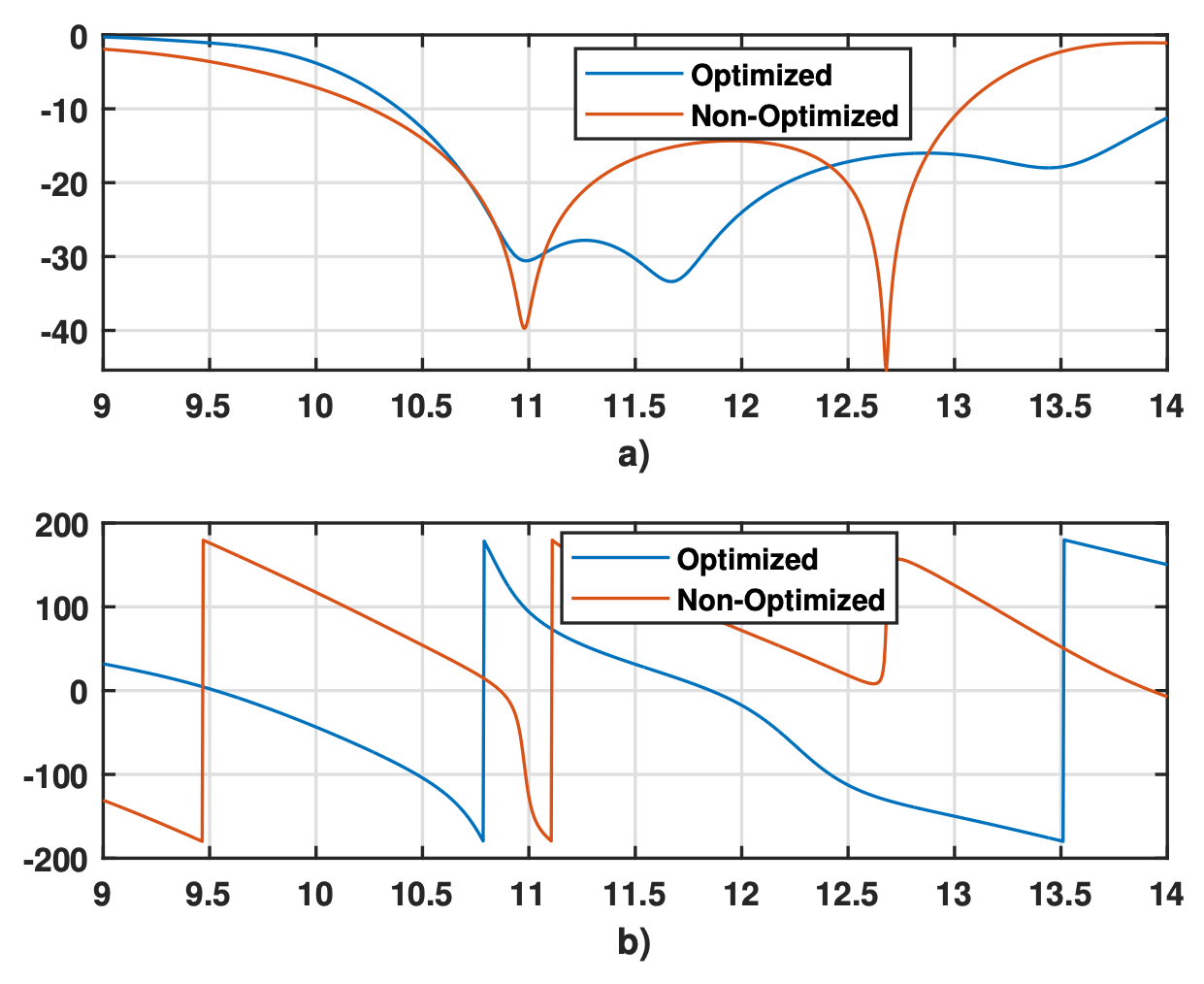}
\caption{Comparison of, a) S-11 parameters after optimization and the non-optimized structure in terms of dB scale, b) phases of optimized and the non-optimized structures in terms of degree.}
\label{fig:Optimization}
\end{figure}

\begin{figure} [!htbp]  
\centering
\includegraphics[width=0.8\columnwidth]
{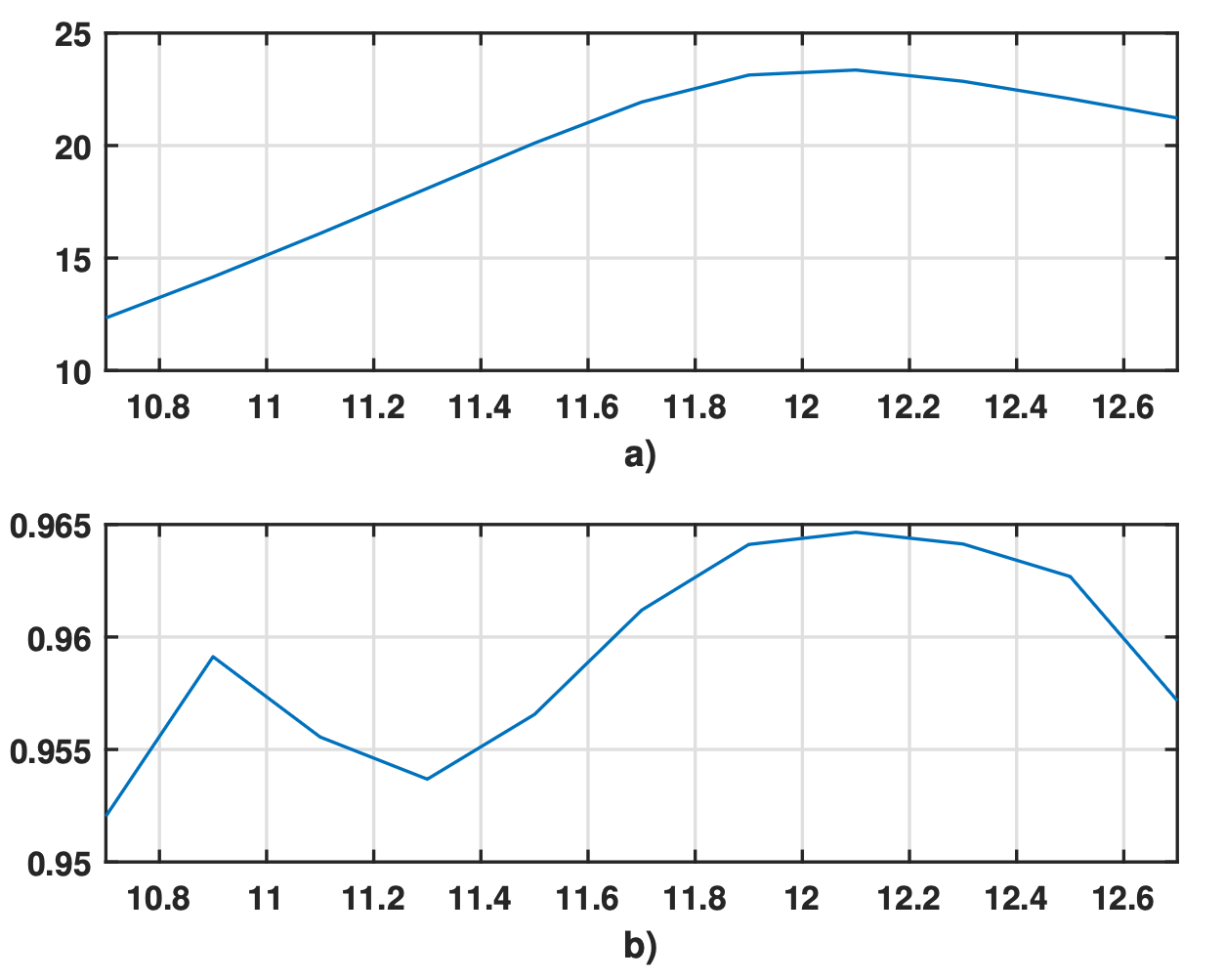}
\caption{Optimized unit cell, a) front back-lobe ratio in dB scale, b) total efficiency in percentage.}
\label{fig:Optimization}
\end{figure}

\subsection{Distribution Network Design}
It is a fundamental component in antenna arrays, facilitating the
distribution of RF signals from a single input to multiple outputs with equal amplitude
and a defined phase relationship. It ensures efficient power split while maintaining impedance
matching, which is crucial for achieving uniform signal distribution among antenna elements in
array configurations. The Wilkinson power divider is a straightforward, cost-effective solution
for distributing signals to multiple antenna elements.

The Wilkinson power divider is a circuit designed to split an input signal into two output signals while maintaining isolation between its output ports, provided all ports are impedance-matched. Each section's length equals one-quarter of the operating wavelength, and the impedance is \( \sqrt{2} \) times the characteristic impedance \( Z_0 \), i.e., \( \sqrt{2} Z_0 \). Additionally, an internal resistor is placed between two separate output ports, with an impedance equal to twice the input characteristic impedance \( Z_0 \), i.e., \( 2Z_0 \). The detailed explanation and design formulations can be found in \cite{pozar_microwave_2021}.

\section{Simulation Results} \label{Sim}
\begin{figure} [!hbt]  
\centering
\includegraphics[width=0.8\columnwidth] {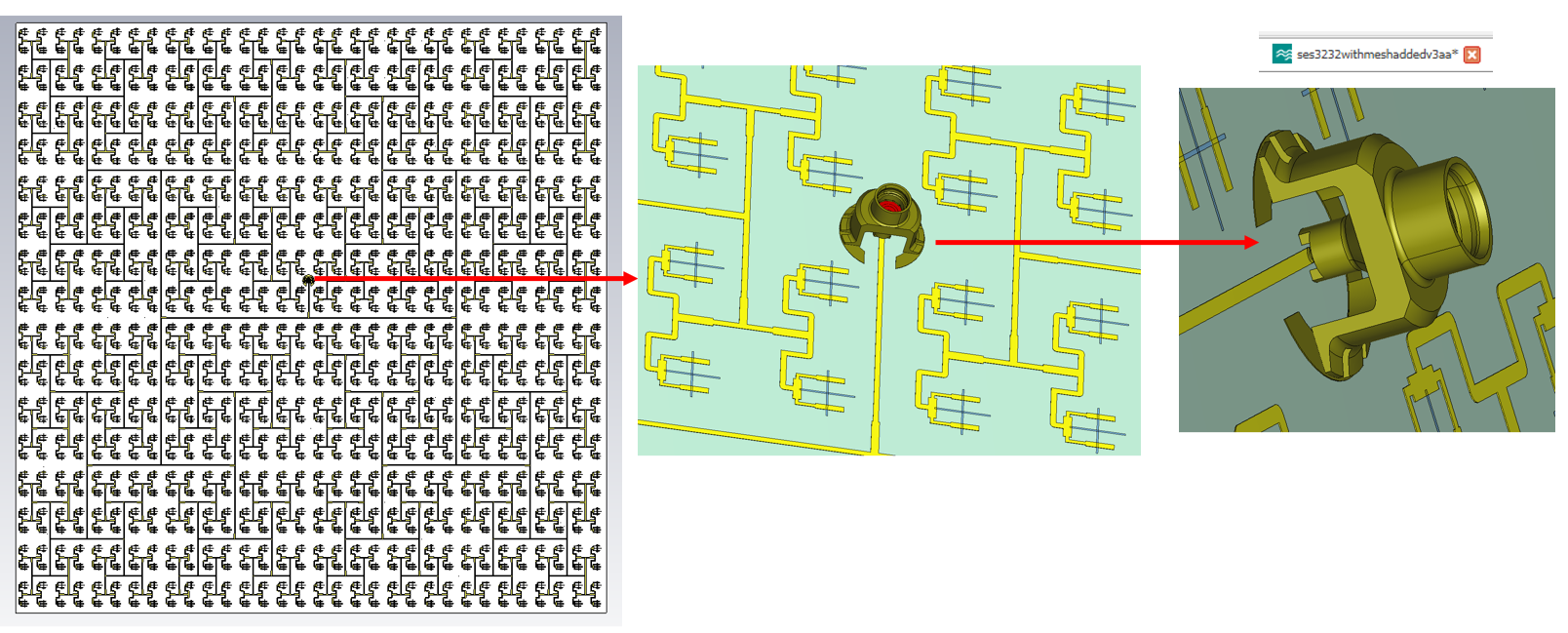}
\caption{32x32 array with surface mount SMP.}
\label{fig:sma1}
\end{figure}

\begin{figure} [!hbt]  
\centering
\includegraphics[width=0.8\columnwidth] {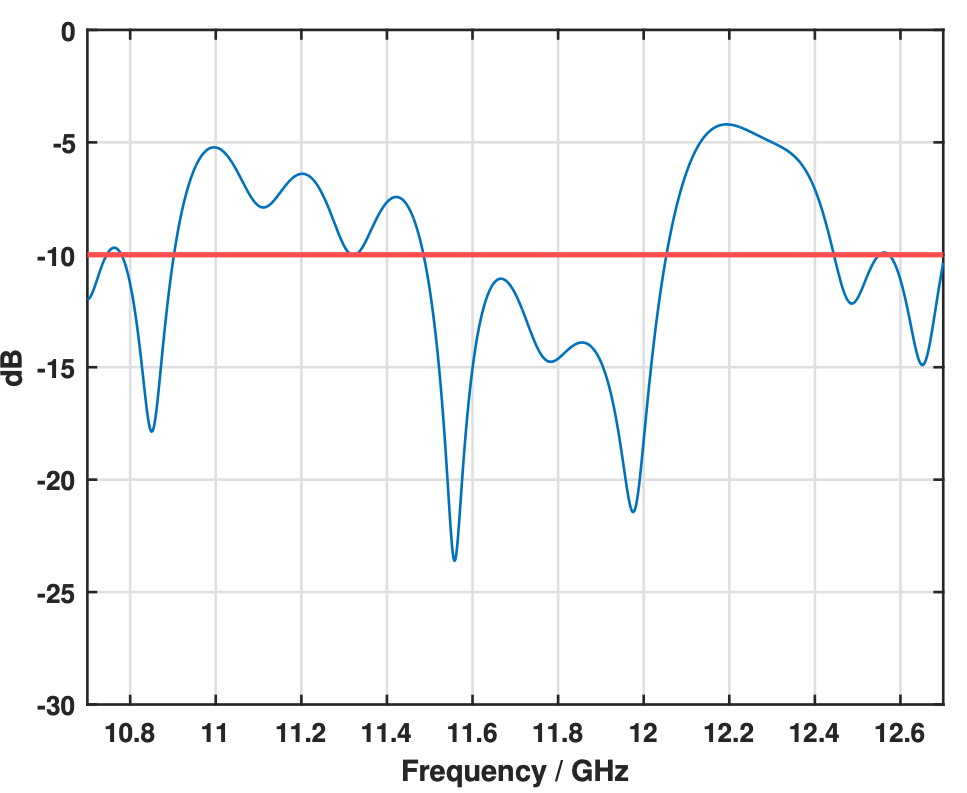}
\caption{Final structure S-11 parameters.}
\label{fig:s11Full}
\end{figure}

The proposed approach involves a symmetrically configured
power divider with antenna unit cells. Initially, we designed a 2x2 element integrated power
divider with edge feeding, leveraging simulations conducted in CST Microwave Studio. Subsequently, the unit cell design was integrated into a 32x32 element
array, with final overall dimensions of 410.26 mm x 410.26 mm.
For the final design, we integrated the surface mount SMP connector \cite{vasquez2021interwoven} at the center of the 32x32 array
configuration, as illustrated in Fig. \ref{fig:sma1}, and analyzed its return loss (S11) characteristics. The
results are presented in Fig. \ref{fig:s11Full}. To delve deeper into these findings, we further investigated the effect of asymmetric and symmetric placement of surface-mount SMP connectors concerning antenna elements. Due to the closely spaced nature of the SMPs and antenna elements, the symmetrical arrangement exhibited poorer return loss performance at higher frequencies compared to asymmetrical arrangements. 

\begin{figure} [!hbt]  
\centering
\includegraphics[width=0.8\columnwidth] {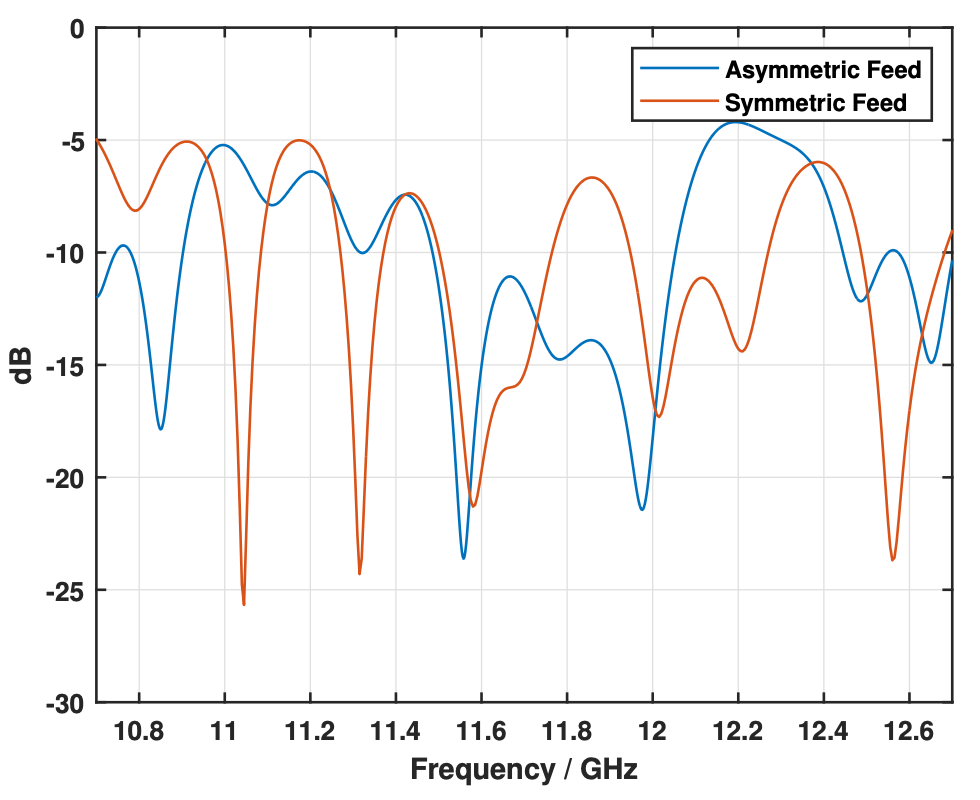}
\caption{S-parameter of 32x32 antenna array with asymmetric and symmetric surface mount SMP position.}
\label{fig:s11FullA}
\end{figure}

To optimize performance, we strategically placed surface-mount SMP connectors asymmetrically in the center of the array. This optimization process included increasing the number of network cells and running detailed simulations to improve performance metrics. 
The compared results are depicted in Fig. \ref{fig:s11FullA}, indicated that the asymmetrical
placement of surface mount SMP connectors significantly improved return loss values, bringing
them within an acceptable range across the desired frequency spectrum.

\FloatBarrier

\begin{figure} [!hbt]  
\centering
\includegraphics[width=0.8\columnwidth] {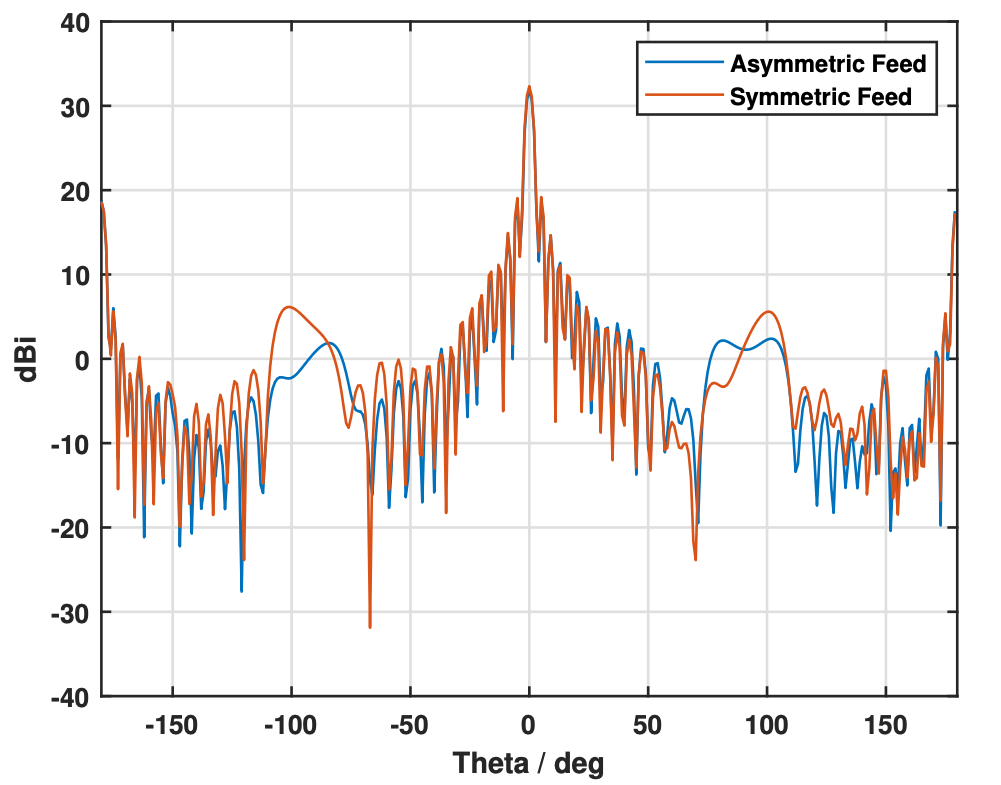}
\caption{Radiation pattern of 32x32 array at 11.7 GHz.}
\label{fig:Pattern1}
\end{figure}

The radiation patterns analysis highlights the presence of side lobe level losses attributed to
back-scattering effects from the power distribution network. These losses indicate areas where
radiation is diverted from the desired main lobe direction, potentially affecting overall
antenna performance. Addressing these challenges involves further optimization of the array
configuration and power distribution strategies to minimize side-lobe effects and enhance overall
radiation efficiency and directivity. The S-parameters of the optimized final asymmetrically fed 32x32 array and its 2D radiation pattern at the center frequency (11.7 GHz) are given in Fig. \ref{fig:s11Full} and Fig. \ref{fig:Pattern1} respectively.

\FloatBarrier

\FloatBarrier

\FloatBarrier

\section{Conclusion} \label{Conc}
The antenna developed in this study exhibits a promising radiation pattern with a main beam achieving more than 30 dBi directivity at the center frequency. Designed with linear polarization, this model lays the foundation for future adaptations to dual-linear and circular polarization (RHCP/LHCP). Although its efficiency is lower than that of parabolic dishes, its slim profile offers significant advantages. Enhancements such as the metasurface layer and a new aperture coupling increase bandwidth capabilities despite the small number of radiating elements. Future developments will focus on integrating the preamplification stage to improve efficiency, using sparse array configurations, and investigating advanced waveguide technologies to reduce distribution losses.

\bibliographystyle{IEEEtran}
\bibliography{referenceAX}

\end{document}